\documentstyle[aps,12pt,epsfig]{revtex}
\begin{document}
\preprint{INJE--TP--98--7}

\title{Potential analysis and absorption cross section in the D1--D5 brane system}

\author{ H. W. Lee and Y. S. Myung }
\address{Department of Physics, Inje University, Kimhae 621-749, Korea} 

\maketitle

\vskip 2in

\begin{abstract}
We analyze the potentials which arise from the D1--D5 brane (5D black hole).
In the sufficiently low energy ($\omega \ll 1$), we can derive the Schr\"odinger-type
equation with potential $V_N$ from the linearized equations.
In this case one can  understand the difference between
 absorption cross section for a free and two fixed
scalars intuitively in terms of their potentials.
In the low temperature limit ($\omega \gg T_H$), one expects the logarithmic
correction to the cross section of a free scalar. However, we cannot obtain
the Schr\"odinger equation with potential for this case.
Finally we comment on the stability of 5D black hole.
\end{abstract}

\newpage
Recently there has been a great progress in the D1--D5 brane system with momentum
along the string direction which  gives us a (D-brane) 5D black hole 
with three charges $(Q_1,Q_5,Q_K)$. The first progress was achieved in the 
Bekenstein-Hawking entropy\cite{Vaf}.
Apart from the success of counting the microstates of a 5D black hole
through D-brane physics, a dynamical 
consideration  becomes an  
important issue\cite{Wad,Das,Cal,Kra}. 
This is so because the absorption cross section (greybody factor)
 for the black hole arises as a 
consequence of the gravitational potential barrier surrounding the 
horizon.  That is, this is an effect of spacetime curvature.  In the effective string 
description, their origin comes from the thermal distribution for
excitations of the D1--D5 bound state.
An effective CFT approach was also introduced to describe  the absorption of scalars
by the general black holes\cite{Mal1}.   
 The 5D black hole becomes 
 $AdS_3\times S^3$  near horizon  but with an asymptotically
flat space\cite{Hyu}. 
In this case the cross section agrees with that for the semiclassical calculation of
 5D black hole\cite{Lee1}. This means that the near horizon geometry contains
the essential information about the bulk 5D black hole.
Also the AdS/CFT correspondence\cite{Mal2} can be used to derive the cross section.
This is so because the $AdS_3\times S^3$  is an exact solution of string theory and there
is an exact CFT on its boundary at spatial infinity.
It turns out that the cross sections in the boundary CFT computation take 
the same forms as those in the semiclassical
and effective string calculations\cite{Teo}.

The calculations of cross section for a minimally coupled scalar 
are straightforward in both semiclassical and effective string models.  
The s-wave cross section is not sensitive 
to the  energy ($\omega$)
but depends only on the area of horizon\cite{Wad,Das}. 
This couples to an operator with dimension (1,1) on the boundary.
A better test of the agreement between semiclassical and effective string  
calculations is provided by the fixed scalars.  The effective string 
calculation is   well performed in  the 
dilute gas limit which corresponds to the decoupling limit.
 But the semiclassical calculations are 
difficult because of a complicated mixing  between fixed 
scalars and other fields (metric and U(1) gauge fields).
One of fixed scalars($\nu$) couples solely to an operator of 
dimension (2,2) on the boundary CFT.  
When $Q_1=Q_5$, 
the effective string calculation of 
yields the precise agreement with the 
semiclassical greybody factor\cite{Cal}.  However,
 the greybody factor of the 
other 
($\lambda$) is not in agreement  for $Q_1=Q_5$\cite{Kra}.  
This disagreement may be  caused by the 
additional chiral operators with dimension (3,1) and (1,3) beyond (2,2) on the boundary. 
This point remains unsolved up to now.

On the other hand, it is  possible to visualize any 
black hole as presenting an effective potential barrier (or well) to the 
on-coming mode\cite{Regge,Chan,Kim1}.   This means that one can derive the Schr\"odinger-type
equation for the physical mode. In this case  one can also  perform
the stability analysis\cite{Myung}.
For example, in  case of the 4D Schwarzschild black hole
\begin{equation}
ds_{4D}^2= - (1 - {r_o \over r}) dt^2 + (1 - {r_o \over r})^{-1} dr^2 + r^2 d\Omega_2^2,
\end{equation}
two graviton modes
arose from the metric perturbations with $l\geq 2$.
One is the Regge-Wheeler (RW) graviton mode in the axial 
(odd-parity) perturbation equation, 
 
\begin{equation}
{d^2 \Psi_{RW} \over d r^{*2} } + (\omega^2 - V_{RW}) \Psi_{RW} = 0. 
\end{equation}
Here a tortoise coordinate $r^* = r + r_o \ln(r-r_o) $ is introduced, so that the horizon is at 
$r^* = - \infty$ ($r=r_o$).
The RW potential $V_{RW}$ is given by
\begin{equation}
 V_{RW} = { 2 (n+1) r - 3 r_o \over r^4 } (r -r_o)
\end{equation}
with  $n = (l -1)(l+2)/2, l\geq 2$.
The other is  the Zerilli mode in the polar(even-parity) equation 
\begin{equation}
{d^2 \Psi_Z \over d r^{*2} } + (\omega^2 -  V_Z) \Psi_Z = 0. 
\end{equation}
which differs only in the details of the potential
\begin{equation}
 V_Z = { 2 (n+1) r^3 + 3 r_o r^2 + 9r r_o^2/2n + 9 r_o^3/4n^2  \over
           r^4 ( r+ 3 r_o/2n)^2  } (r -r_o).   
\end{equation}
Although these have different forms,
Chandrasekhar have showed that $V_{RW}$ and $V_Z$ are equivalent in the sense of producing 
the same reflection $({\cal R})$ and 
absorption $({\cal A})$ coefficients\cite{Chan}.
For a mimimally coupled scalar $(\psi)$, one finds
\begin{equation}
{d^2 \psi \over d r^{*2} } + (\omega^2 - V_\psi) \psi = 0, 
\end{equation}
where the potential is given by
\begin{equation}
 V_\psi ={ 2 (n+1) r + r_o \over r^4 } (r -r_o)
\end{equation}
with  $l\geq 0$\cite{Myung}. 
As is shown in Fig. 1, the lowest allowed potentials with $r_o = 0.01$ take all barrier-type.
Here $n=2(l=2)$ for RW, Zerilli modes and $n=-1(l=0)$ for $\psi$.
$V_{RW} \simeq V_Z$ implies the same reflection and absorption 
coefficients.
The s-wave absorption cross section for a free scalar $\psi$ 
is $\sigma_{4D}^\psi = A_H^{4D}= 4 \pi r_o^2$.
\begin{figure}
\epsfig{file=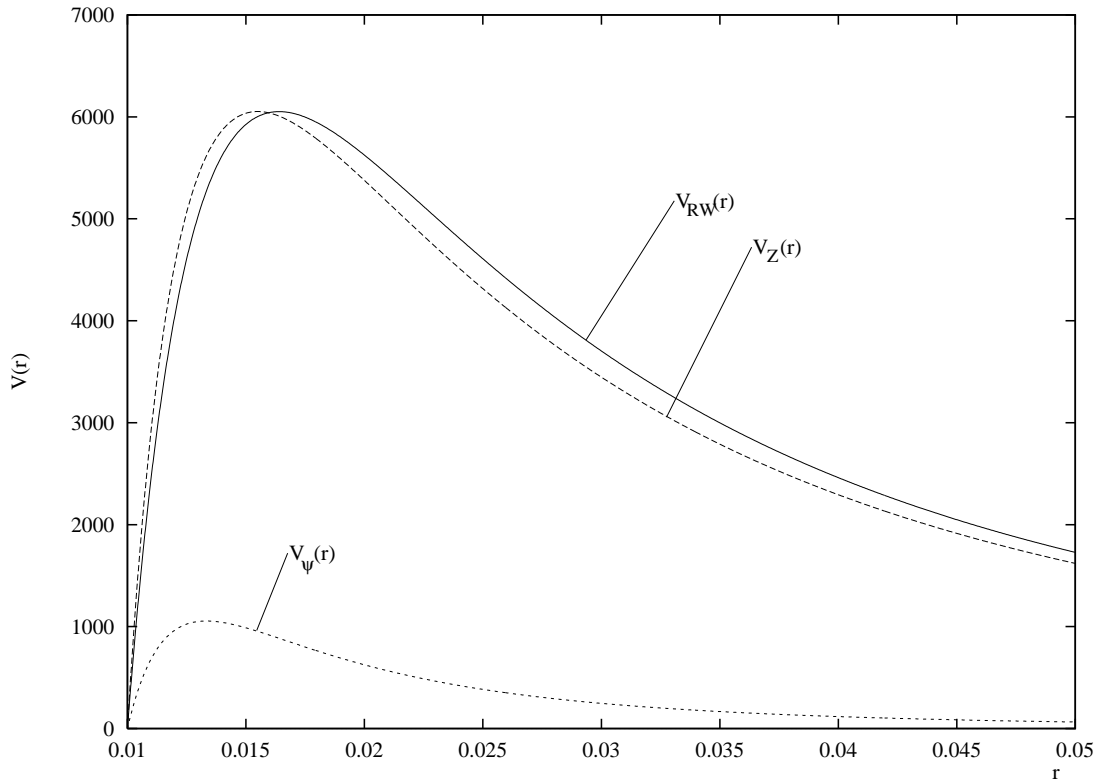,width=0.9\textwidth,clip=}
\caption{Three potential graphs ($V_{RW}, V_Z, V_\psi$) for 4D 
Schwarzchild black hole with $r_0=0.01$.}
\end{figure}

In this paper we will clarify the close relationship between the potential and absorption 
cross section in the D1--D5 brane system. 
Initially we introduce  all modes around the 5D black hole background.
It is pointed out that in s-wave calculation  fixed scalars are 
physically propagating modes and other fields belong to redundant modes.  
The relevant modes are two fixed scalars $(\nu,\lambda)$ including a free scalar$(\phi)$.
We begin with the 5D black hole with three charges,
\begin{equation}
ds_{5D}^2= - h f^{-2/3}dt^2 +  f^{1/3}( h^{-1} dr^2 + r^2 d\Omega_3^2),
\end{equation}
where
\begin{equation}
f = f_1 f_2 f_K = (1 + { r^2_1 \over r^2})(1 + { r^2_5 \over r^2})(1 + { r^2_K \over r^2}),~~~
h = (1 - { r^2_0 \over r^2}).
\end{equation}
Here the radii are related to the boost parameters $(\alpha_i)$ and the charges $(Q_i)$ as
\begin{equation}
 r_i^2 = r_0^2 \sinh^2 \alpha_i = \sqrt{Q_i^2 + {r_0^4 \over 4}} - 
{r_0^2 \over 2}, i = 1, 5, K.
\end{equation}
Hence the D-brane black hole depends on the four parameters $(r_1, r_5, r_K, r_0)$.
The background metric (8) is just the 5D Schwarzschild one 
with time and space components rescaled by different powers of $f$.  The event 
horizon (outer horizon) is clearly at $r=r_0$.  When all three charges are 
nonzero, the surface $r=0$ becomes a smooth inner horizon (Cauchy horizon).  
When at least one of the charges is zero, 
the surface $r=0$ becomes singular.  The 
extremal case corresponds  to the limit of $r_0 \rightarrow 0$ with the boost 
parameters $\alpha_i \rightarrow \pm \infty$, keeping the  charges 
$(Q_i)$ fixed.   We are interested in the limit of
 $r_0, r_K \ll r_1, r_5$, which is called 
the dilute gas region. Here we have $Q_1 = r_1^2, Q_5 = r_5^2$, and $r_K = r_0 \sinh \alpha_K$
 with a finite $\alpha_K$. This corresponds to the near extremal black hole 
and its thermodynamic quantities (energy, entropy, Hawking temperature) are given by
\begin{eqnarray}
&&E_{next} = {2 \pi^2 \over \kappa_5^2} \left [ r_1^2 +
r_5^2 + {1 \over 2}r_0^2\cosh 2 \alpha_K \right ],\\
&&S_{next} = { 4 \pi^3 r_0 \over \kappa_5^2} r_1 r_5\cosh \alpha_K,\\
&&{1 \over T_{H,next}} = {2 \pi \over r_0} r_1 r_5 \cosh \alpha_K,
\end{eqnarray}
where $\kappa^2_5$ is the 5D gravitational constant. 
The above energy and entropy are those of a gas of massless 1D particles.  
In this case the  temperatures for left and right moving
string modes are given by
\begin{equation}
T_L = {1 \over 2 \pi} \left ( {r_0 \over r_1 r_5} \right ) e^{\alpha_K},~~
T_R = {1 \over 2 \pi} \left ( {r_0 \over r_1 r_5} \right ) e^{-\alpha_K}.
\end{equation}
This implies that the (left and right moving) momentum modes along the string direction  are 
excited, while the excitations of D1--anti D1 and D5--anti D5-branes
are suppressed.  The Hawking temperature is given by 
their harmonic average
\begin{equation}
{2 \over T_H} = {1 \over T_L} + {1 \over T_R}.
\end{equation}
We take here  $r_1 = r_5 = R$ and $r_0 = r_K$ for simplicity. Then the linearized equation
for s-wave fixed scalar take the form\cite{Cal,Kra}
\begin{equation}
\left [ ( h r^3 \partial_r)^2 + \omega^2 r^6 f - 
  {{8 h r^4 r_{\pm}^4} \over (r^2 + r_{\pm}^2)^2 }
  \left ( 1 + {r_0^2 \over r_{\pm}^2} \right ) \right ] \tilde \phi_{\pm}=0,
\end{equation}
where one gets $\delta \tilde \nu$, for $r_+^2 = R^2$ and $\delta \tilde \lambda$,
 for $r_-^2 = R^2/3$.
For a minimally coupled scalar $(\phi)$, the equation leads to\cite{Mal1}
\begin{equation}
\left [ ( h r^3 \partial_r)^2 + \omega^2 r^6 f - 
  {  l(l+2) h\over r^2  }
   \right ] \delta \tilde \phi = 0.
\end{equation}
Considering $\delta \tilde N = r^{-3/2} \delta N$, for $N = \nu, \lambda, \phi$ and 
introducing a tortoise coordinate $ r^* = \int {(dr/h)} = r + (r_0/2)\ln |(r - r_o)
/(r + r_0)|$\cite{Wad}, then the equation takes the form
\begin{equation}
{d^2 \delta N \over d r^{*2} } + (\omega^2 - \tilde V_N) \delta N = 0. 
\end{equation}
Here $\tilde V_N$ in the dilute gas limit is given by
\begin{equation}
\tilde V_\nu = - \omega^2 (f -1)  
+ { 3  h  \over 4 r^2 } (1 + { 3 r_0^2 \over r^2} ) + { 8 R^4 h  \over r^2 (r^2 + R^2)^2 },
\end{equation}
\begin{equation}
\tilde V_\lambda = - \omega^2 (f -1)   
+ { 3  h  \over 4 r^2 } (1 + { 3 r_0^2 \over r^2} ) + { 8 R^4 h  \over r^2 (3 r^2 + R^2)^2 },
\end{equation}
\begin{equation}
\tilde V_\phi = - \omega^2 (f -1)  
+ { 3  h  \over 4 r^2 } (1 + { 3 r_0^2 \over r^2} ) + { l(l+2) h  \over r^2  },
\end{equation}
where 
\begin{equation}
f -1 = { r_0^2 + 2 R^2 \over r^2} + { (2 r^2_o + R^2) R^2 \over r^4}
        + { r^2_0 R^4 \over r^6}.
\end{equation}

\begin{figure}
\epsfig{file=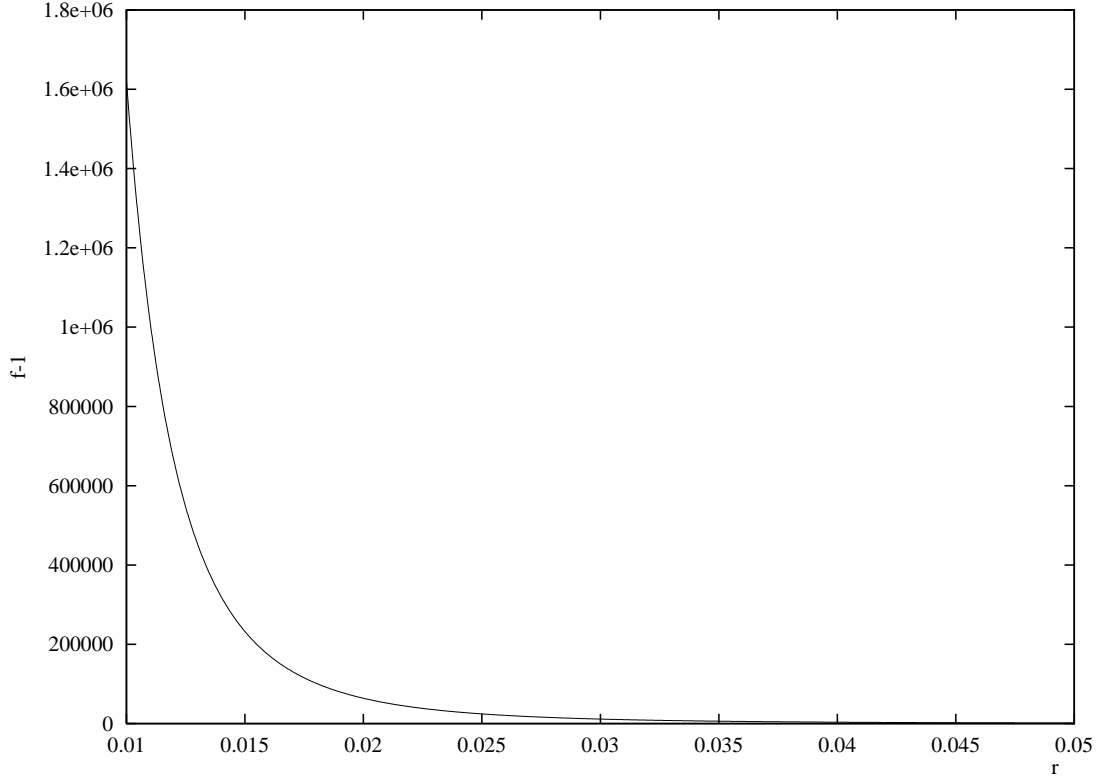,width=0.9\textwidth,clip=}
\caption{The graph of $(f-1)$ in 5D black hole with $r_0=0.01, R=0.3$. 
A peak appears near horizon($r=r_0$).}
\end{figure}

We note that $\tilde V_N$ depends on two parameters ($r_0, R$)
as well as the energy$(\omega)$. As (18) stands, it cannot be considered as
the Schr\"odinger equation. The $\omega$-dependence is a matter of peculiar interest
to us compared with the 4D black hole potentials $(V_{RW}, V_Z, V_\psi)$. This makes the 
interpretation of $\tilde V_N$ as a potential difficult. 
As is shown in Fig. 2, this is so because $(f-1)$ is very large as $10^6$ for $r_0 = 0.01,
R= 0.3 $ near horizon.
In order for $\tilde V_N$ to be a potential, 
it is necessary to take the sufficiently low energy limit of $\omega \to 0$.
It is suitable to be $10^{-3}$.  
And $\omega^2(f-1)$ is of order ${\cal O}(1)$ and thus
it can be ignored in comparison to the remaing ones. 
Hence we  define  a potential $V_N$ to be $\tilde V_N$ without $\omega^2(f-1)$. 
Further the last terms in (19)-(21) are important to compare each other.
After the partial fraction, the last terms in (19)-(20) lead to
\begin{eqnarray}
&& { 8 R^4   \over r^2 (r^2 + R^2)^2 }= { 8 \over r^2} - { 8 \over r^2 + R^2}
- { 8 R^2 \over (r^2 + R^2)^2},\label{par-far1}\\
&&{ 8 R^4   \over r^2 (3 r^2 + R^2)^2 }= { 8 \over r^2} - { 24 \over 3 r^2 + R^2}
- { 24 R^2 \over (3 r^2 + R^2)^2}.
\end{eqnarray}
The last term in (21) for a minimally coupled scalar with $l=2$ keeps the first
terms  in (23) -(24). Thus one finds immediately the sequence 
\begin{equation}
 V_{\phi_0} \ll V_\lambda \leq V_\nu \leq V_{\phi_2}.
\end{equation}
Here the subscript $\phi_0$ denotes the s-wave($l=0$) free scalar and
 $\phi_2$ the  free one with $l=2$.
This is confirmed from the graphs of potential in Fig.3 
with $ r_0 = 0.01, R = 0.3$.
It is conjectured that $V_\lambda \simeq V_\nu \simeq V_\phi^{l=2}$ 
gives us the nearly same ${\cal R}$ and ${\cal A}$.  This 
implies the nearly same absorption cross section because of 
$\sigma_{5D} = 4 \pi {\cal A} /\omega^3$.
\begin{figure}
\epsfig{file=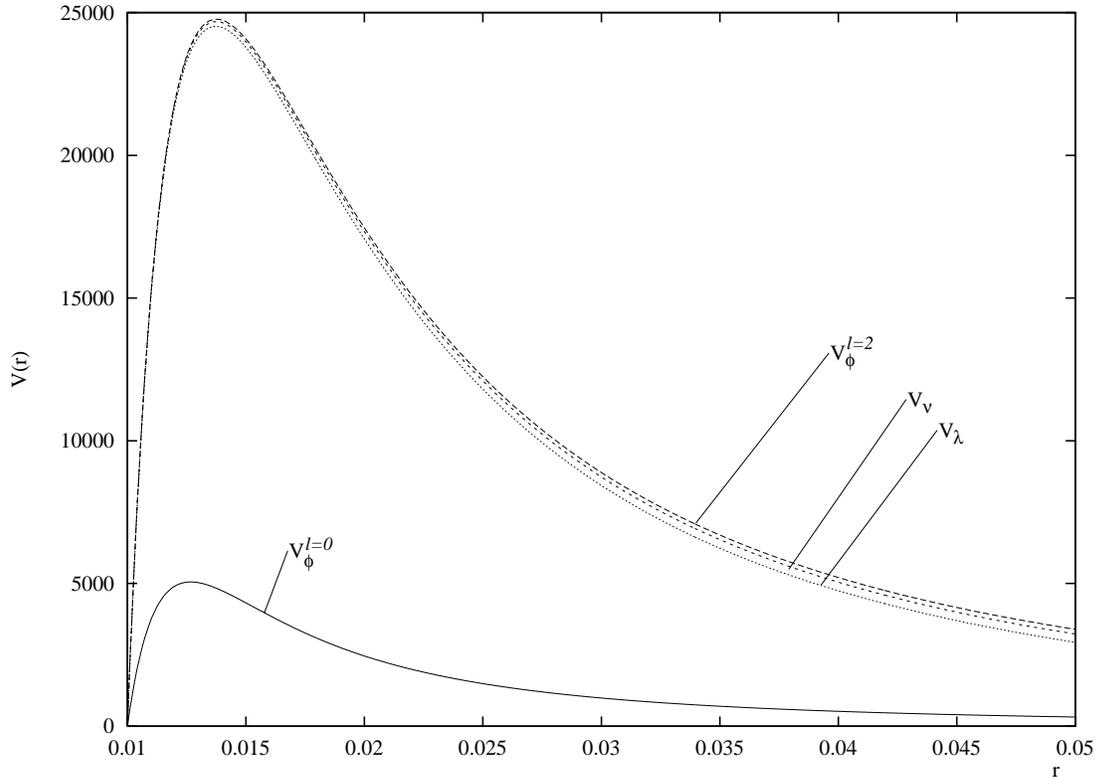,width=0.9\textwidth,clip=}
\caption{ 
Four potential graphs ($V_\phi^{l=0}, V_\nu, V_\lambda, V_\phi^{l=2}$) 
for 5D black hole with $r_0=0.01, R=0.3$.}
\end{figure}

On the other hand, using Eqs.(16)-(17), the low-energy absorption cross sections are
calculated as\cite{Kra,Lee1}
\begin{eqnarray}
&& \sigma_{5D}^{\phi_0}  = A_H^{5D},\\
&&\sigma_{5D}^{\phi_2}  = {3 \over 16} (\omega r_0)^4 = {3 \over 4} (\omega R)^4
             {A_H^{5D} \over 4}({r_0 \over R})^4 ,\\
&& \sigma_{5D}^{\nu}  = {A_H^{5D} \over 4}({r_0 \over R})^4,\\
&&\sigma_{5D}^{\lambda}  = 9{ A_H^{5D} \over 4}({r_0 \over R})^4,\\
\end{eqnarray}
with the area of horizon $A_H^{5D} = 2 \pi^2 R^2 r_K$ for the 5D black hole.
In deriving the above, one uses the condition of $\omega < T_L, T_R, T_H$.
Here we find a sequence of cross section
\begin{equation}
\sigma_{5D}^{\phi_0} \gg \sigma_{5D}^{\lambda} \geq \sigma_{5D}^{\nu} \geq \sigma_{5D}^{\phi_2}.
\end{equation} 
This originates from the potential sequence in (25).
It is consistent with our naive expectation that
the absorption cross section increases, as the height of potential decreases.
Here we wish to point out the difference between a free and fixed scalar.
In the dilute gas limit ($ R \gg r_0$) and the low energy  limit ($\omega R \ll1$),
 the s-wave cross section for a minimally coupled scalar($\sigma_{5D}^{\phi_0}$) 
goes to $A_H^{5D}$\cite{Das},
 while the s-wave cross sections for fixed scalars ($\nu,\lambda$) 
including $\phi_2$ approach zero\cite{Cal}. 
This is consistent with our conjecture from (25).

Now we are in a position to discuss the $AdS_3\times S^3$-theory.
In the near horizon, we approximate $(f -1)$ in (19)-(21) as 
\begin{equation}
(f -1) \approx (f-1)^{AdS} = { R^4 \over r^4} \big ( 1 + { r_0^2 \over r^2} \big ).
\end{equation}
while the other terms remain invariant.
The potential ($V_N^{AdS}$) where $f-1$ is replaced by $(f-1)^{AdS}$ 
corresponds to that for the $AdS_3\times S^3$-theory.
In the sufficiently low energy limit of $\omega \sim 10^{-3}$,
 two potential $(V_N, V_N^{AdS})$ take
 the same form. Thus we expect that two cross sections are same.
Actually it turns out that the cross sections for the $AdS$-theory take the same form
as (26) and (28) \cite{Lee1}
\begin{eqnarray}
&& \sigma_{AdS}^{\phi_0}  = A_H^{6D},\\
&&\sigma_{AdS}^{\phi_2}  = \sigma_{AdS}^{\nu}  = 
{1 \over 3}{A_H^{6D} \over 4}({r_0 \over R})^4
\end{eqnarray}
with the area of horizon $A_H^{6D}=A_H^{5D} \times 2 \pi R = 4 \pi^3 R^3 r_K$
 for the $AdS_3 \times S^3$.
We note that $\phi_2$ and $\nu$ give us  slightly different cross sections in the D-brane
black hole, whereas these do not make any distinction in the $AdS$-theory.

However, in the low temperature limit $( \omega \gg T_H, T_L, T_R)$, $\omega^2(f-1)$-term
plays an important role. Here we  have to assume the low energy scattering with
$\omega R \ll 1$. For example we choose $\omega^2 \sim 10^{-3}$ and
 $\omega^2(f-1) \sim 10^3$. Then this becomes comparable with $V_{\phi_0}$.
This can be observed from the behavior of $f -1$ in Fig. 2 and $V_{\phi_0}$ in Fig. 3.
 In this case one expects the cross section to behave as\cite{Rob}
\begin{equation}
\tilde \sigma_{5D}^{\phi_0}  = A_H^{5D}[ 1 + {\cal O} (\omega R)^2 \ln(\omega R)].
\end{equation}
The logarithmic correction term encodes the leading order departure from the conformal limit.
This means that nonrenormalizable interactions
enter into the world sheet action at the subleading order.
 However, in the semiclassical approach, this implies that
 we cannot obtain the Schr\"odinger equation with potential.
This is because the $\omega$-dependence term ($\omega^2(f-1)$) is included as a part of
the potential and is comparable with $V_{\phi_0}$.

In conclusion, we analyze the physical potentials surrounding the D-brane black hole.
There is an essential difference between a free and fixed scalars. But the distinction
between two fixed scalars $(\nu, \lambda)$ is not clearly understood in the semiclassical
approach.
We express the difference between
 absorption cross section for a free and two fixed
scalars  in terms of their potentials.
 As in the 4D Schwarzschild black hole,
 they take all barrier-types in the low energy limit.
This implies that there is no exponentially growing mode 
and thus this black hole
is stable against the s-mode perturbations\cite{Regge,Chan,Kim1,Myung}.
Here we note that the stability analysis should be based on the physical modes.
In our case these are two fixed scalars. 

In the low temperature limit, we cannot obtain the Schr\"odinger-type equation.
Thus  the meaning of potential is unclear and the analysis of stability  is obscure.
Also this is related to  the logarithmic correction of cross
section.
It seems that there is a relation between the potential $(V_N^{AdS})$
and conformal symmetry on the boundary at the spatial infinity\cite{Bal}.
We expect that this can be understood from the AdS/CFT correspondence\cite{Car,Boe}.
\acknowledgments
This work was supported in part by the Basic Science Research Institute 
Program, Ministry of Education, Project NO. BSRI--98--2413.

\end{document}